\documentclass{llncs}
\usepackage{makeidx}  % allows for indexgeneration
\usepackage{amssymb}
\usepackage{graphicx}
\usepackage{comment} 
\usepackage{multirow}

\usepackage{url}
\urldef{\mails}\path|{jing.liu, john.backes, darren.cofer, andrew.gacek}@rockwellcollins.com|    
\newcommand{\keywords}[1]{\par\addvspace\baselineskip
\noindent\keywordname\enspace\ignorespaces#1}

\begin{document}

\mainmatter

\title{From Design Contracts to Component Requirements Verification}

\author{Jing (Janet) Liu
\and John D. Backes \and Darren Cofer\and Andrew Gacek }

\institute{Advanced Technology Center, Rockwell Collins\\
%400 Collins Rd. NE, Cedar Rapids, IA, 52498, USA\\
\mails\\}

\maketitle

\begin{abstract}
During the development and verification of complex airborne systems, a variety
of languages and development environments are used for different levels of the
system hierarchy. As a result, there may be manual steps to translate
requirements between these different environments. This paper presents a
tool-supported export technique that translates high-level requirements from the
software architecture modeling environment into observers of requirements that can be
used for verification in the software component environment. This allows
efficient verification that the component designs comply with their high-level
requirements.  It also provides an automated tool chain supporting formal
verification from system requirements down to low-level software requirements
that is consistent with certification guidance for avionics systems.
The effectiveness of the technique has been evaluated and demonstrated on a
medical infusion pump and an aircraft wheel braking system.

%\keywords{Design contracts, component requirements, AADL, Simulink}
%\keywords{design contracts, verification cases, requirements-based verificaiton, compositional analysis}
\keywords{design contracts, specification model, design model, AGREE, Simulink, requirements-based verification, certification}
\end{abstract}

\section{Introduction} \label{sec:intro}

As part of the software development process for complex avionics systems, system
requirements are iteratively decomposed, allocated, and refined to lower level
requirements for software and hardware components. Different verification
processes are used to provide evidence that these components satisfy their
requirements. The focus of all development and verification activities in the
avionics domain is to ensure that a system meets its requirements and contains
no unintended functionality.

Requirements at different levels of the system hierarchy may be specified using
different languages and development environments. Even when formal methods tools
are used to verify requirements, there may be manual steps to translate
requirements between these different environments. The work presented in this
paper attempts to close the gap between verification at the system level and the
component level.

We present a tool-supported technique that translates requirements from a system-level
reasoning framework into observers of requirements for software components. The
observers that the tool produces can be verified using a model checker
specialized to the software component development language. Our work closes the
gap between high-level requirements captured with the software architecture and
low-level requirements for component implementation. This ensures consistency of
the verification results and improves productivity and accuracy through the use
of automation to eliminate manual steps. Furthermore, making these property
observers available during the design process supports early verification of
the software components.

In previous work, we have developed a compositional analysis environment
\cite{Cofer12:comp-verif,Whalen13:what-is-how} based on the Architecture
Analysis and Design Language (AADL) \cite{Feiler12:AADL-MBE-book}. AADL can be
used to model both the hardware and software aspects of the system, but in this
work we have limited our attention to the software architecture. In our
compositional analysis approach, the AADL model is augmented with
assume-guarantee contracts to capture both system-level requirements and the
requirements for the software components.

In the present work, we link the component contracts to their implementations
in Simulink\textsuperscript{\textregistered}~\cite{Simulink}, a framework developed by MathWorks\textsuperscript{\textregistered} and integrated
with MATLAB\textsuperscript{\textregistered}. Simulink provides a graphical programming environment for modeling,
simulation, code generation, testing, and formal analysis. It is widely used in
the avionics industry. By automatically translating formal contracts for
software component behavior into specifications that can be checked in the
Simulink environment, we now support a complete top-to-bottom development
process with formal verification of all requirements. Furthermore, the design of
our approach is sufficiently general that it can be adapted to support other
software development environments and languages.

Since our objective is the production of high-assurance software for avionics,
we must be cognizant of how this approach will fit into a certification context.
As we will show, our approach has been designed to be consistent with new
certification guidance related to the use of formal methods and model-based
development processes.

The rest of the paper is organized as follows. Section \ref{sec:prelims}
provides background information related to the development and analysis
environment, including certification considerations. Section
\ref{sec:detailed_approach} describes the contract translation process in
detail. Section \ref{sec:case_study} evaluates the techniques in an avionics
system case study and a medical device system case study. Section
\ref{sec:related_work} describes related work and Section
\ref{sec:conclusion} presents concluding remarks.

\section{Preliminaries}
\label{sec:prelims}

In this section, we describe the overall design flow and introduce
some terminology associated with the certification context. We also
describe the system architecture modeling environment and the software
component modeling environment that we are using.

%\section{Certification Aspects}
\subsection{Design Flow from Architecture to Component}
\label{subsec:cert}

One of our goals is to transition the tools we have developed into use by the system and software engineers who develop avionics products. 
Therefore, we need to understand how the tools and the models they produce will fit into the certification process.  

Certification guidance for software in commercial aircraft is found in DO-178C, {\em Software Considerations in Airborne Systems and Equipment Certification} \cite{DO178C}.  The process described in DO-178C is essentially a waterfall model in which system requirements are allocated to hardware and software, becoming high-level requirements for each.  High-level requirements are refined to become a software design, consisting of software architecture and low-level requirements from which individual software components can be developed.  

DO-178C is accompanied by several supplement documents which provide guidance for the use of specific technologies, including formal methods (DO-333 \cite{DO-333}) and model-based development (DO-331 \cite{DO-331}).  DO-333 describes how software life-cycle artifacts such as high and low-level requirements can be expressed as formal properties and how formal analysis tools can be used to satisfy many certification objectives.  DO-331 provides guidance on how software life-cycle artifacts expressed as different types of models fit into the certification process.  A case study showing how different formal methods can be used to satisfy certification objectives is found in \cite{casestudy}, including a model-based example that uses Simulink \cite{Simulink} and Simulink Design Verifier\textsuperscript{TM} \cite{SLDV}.  

DO-331 describes the relationships between models at the system and software levels, and distinguishes between {\em specification models} and {\em design models}. A specification model represents high-level requirements that provide an abstract representation of functional, performance, interface, or safety characteristics of software components.  Specification models do not define software design details or prescribe a specific software implementation or architecture.  Design models prescribe software component internal data structures, data flow, and/or control flow.  They may include low-level requirements or architecture, and may be used to produce source code directly.  

\begin{figure}[t]
	\begin{center}
		\includegraphics [height=2.5in] {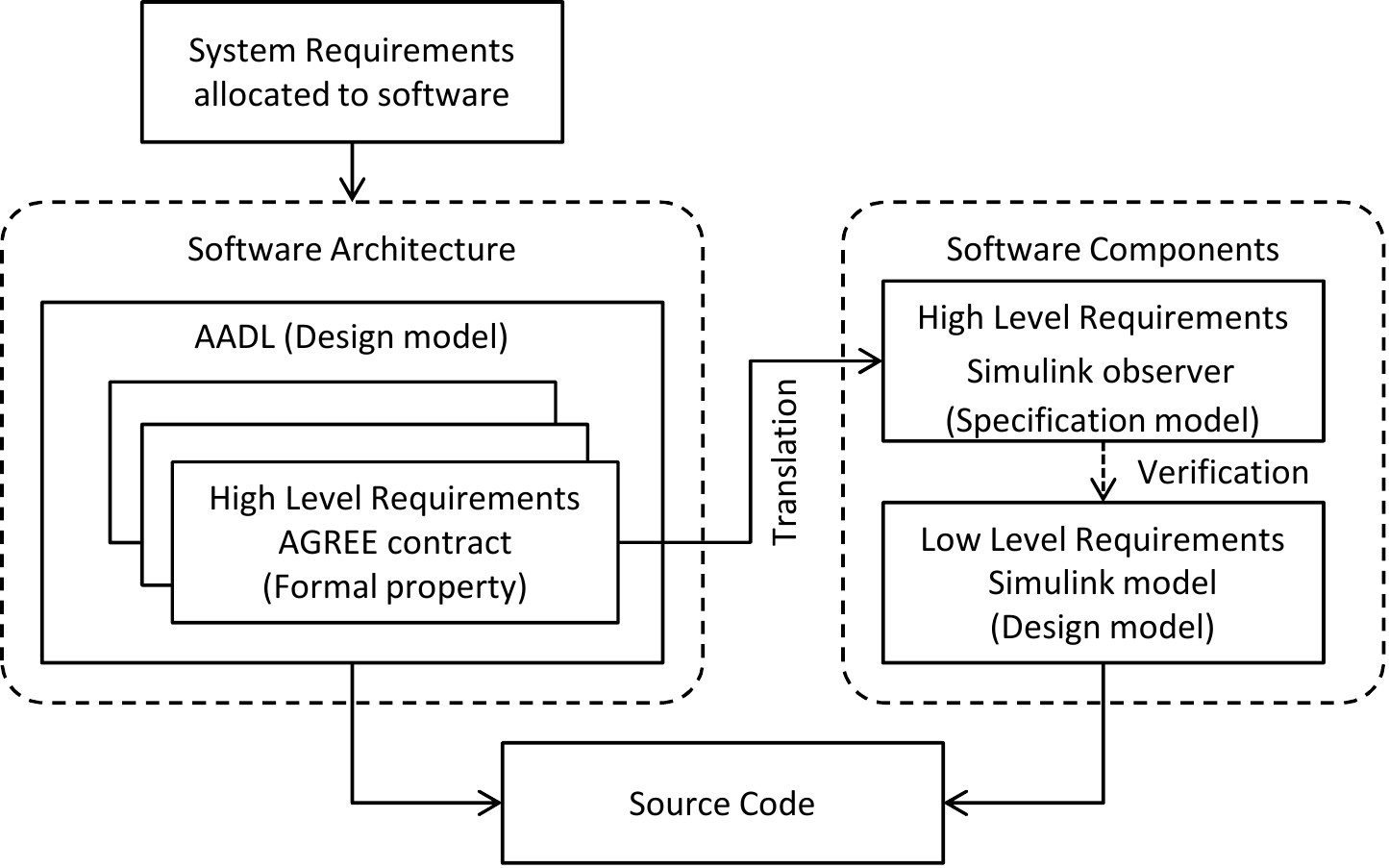}
	\end{center}
	\caption{Architecture to Component Design Flow}
	\label{fig:design-flow}
\end{figure}

Figure~\ref{fig:design-flow} provides an overview of our proposed design flow, connecting it to the terminology used in a DO-178C process.  
%What we have in mind is most like Example A in DO-331 section MB-B.17.  In that example, system requirements allocated to software are used to guide development of a software architecture design model and software component design models, also considered to be low-level requirements for the software.  The addition of formal verification makes our approach somewhat different from the example.  
On the left side of Figure~\ref{fig:design-flow}, system requirements
allocated to software (generally in textual form) are refined to a
collection of high-level software requirements and used to construct
an architecture model in AADL. This process is described in more
detail in the next section. The AADL model is a design model (in
DO-331 terminology) because it contains information such as data
flows, message types, and execution rates and priorities, that will be
used to produce source code and configure the operating system.
High-level requirements associated with each level of the architecture and software
components represented in the architecture are captured into formal
design contracts using the Assume Guarantee Reasoning Environment (AGREE) \cite{Cofer12:comp-verif}. Compositional
verification is used to show that contracts (requirements) at each
level satisfy the contract of the level above.

\begin{comment}
High-level requirements associated with each level of the architecture and software
components represented in the architecture are captured into formal
properties. We use the Assume Guarantee Reasoning Environment (AGREE) \cite{Cofer12:comp-verif}
to represent these properties as design contracts, and compositional
verification is used to show that contracts (requirements) at each
level satisfy the contract of the level above.
\end{comment}

On the right side of the figure, software components are implemented and verified. Simulink models describe the detailed behaviors and are used to generate source code for each component.  They are therefore considered low-level requirements and also design models (in DO-331 terms).  High-level requirements for each component are represented as specification models (in DO-331 terms).  These models are observers that produce a true output whenever their corresponding property (specified over the component inputs and outputs) is true.  A model checker such as the one provided by Simulink Design Verifier can be used to show that the design model satisfied the high-level requirements defined by the specification model.  

Clearly there is a gap between the methods, tools, and models of the software architecture and those for the software components.  In the past, high-level requirements for the software components have been manually captured as observers in Simulink before they can be used for verification \cite{Murugesan:2013:CVM:GPCA}.  The manual process may be error-prone, and it can be difficult and costly to keep the models in sync. The work we describe in this paper bridges this gap by automating the translation of high-level requirements associated with the architecture model into Simulink observers that can be verified in the Simulink environment.

\subsection{Architecture Description and Design Contracts}
\label{subsec:aadl-agree}

The Architectural Analysis and Design Language (AADL)
\cite{Feiler12:AADL-MBE-book} is a architecture modeling language for embedded,
real-time, distributed systems. It was approved as an SAE Standard in 2004, and
its standardization committee has active participation from many academic and
industrial partners in the aerospace industry. It provides the constructs needed
to model both hardware and software in embedded systems such as threads, processes, processors, buses, and memory. It is sufficiently formal for our purposes,
and is extensible through the use of language annexes that can initiate calls to
separately developed analysis tools.

\begin{comment}
A model described in AADL consists of
a number of components and connections. Components are typical
constructs that appear in embedded systems. Components can represent
hardware constructs (e.g., processors, buses, memories) or software
constructs (e.g., processes, threads, subprograms). The language
distinguishes between component types and implementations. A component
type describes the component's interface (e.g., inputs, outputs) as
well as other component specific properties. A component
implementation describes the subcomponents and connections that
implement this interface.
\end{comment}

The Assume Guarantee Reasoning
Environment (AGREE) \cite{Cofer12:comp-verif} 
is a language and tool for compositional verification of AADL models.  It is implemented as an AADL annex
that allows AADL models to be annotated with  assume-guarantee behavioral contracts.  
A contract contains a set of assumptions about the component's
inputs and a set of guarantees about the component's outputs. The assumptions
and guarantees may also contain predicates that reason about how the state of
a component evolves over time.

AGREE uses a syntax similar to Lustre \cite{Lustre-91} to express a contract's assumptions
and guarantees. AGREE translates an AADL model and its contract annotations
into Lustre and then queries a user-selected model checker to perform verification.
The goal of the analysis is to prove that each component's contract
is satisfied by the interaction of its direct subcomponents as
described by their respective contracts. Verification is performed at each
layer of the architecture hierarchy and details of lower level
components are abstracted away during verification of higher level
component contracts.  This compositional approach allows large systems to be analyzed efficiently.  

Component contracts at the lowest level of the
architecture are assumed to be true by AGREE. Verification of these
component contracts must be performed outside of the AADL/AGREE environment.  
In a traditional software development process, components will be developed to meet
the high-level requirements corresponding to these contracts and verified by testing or code review.
However, there are two problems with this approach:

\begin{enumerate}
\item Verification methodologies like test and code review are not exhaustive. 
Errors in these activities can cause the compositional verification that 
AGREE performs to be incorrect.
\item Manual translation of an AGREE contract into a property for verification at the component level can be
time-consuming and error-prone.
%There may be errors in the translation of the AGREE contract from the 
%properties that were proved from component level verification activities.
\end{enumerate}

Our solution to these problems is to automatically translate AGREE
contracts of software components into
expressions in the development language of the component software. A formal
verification tool that reasons about artifacts expressed in this
language can then be used to verify that the contracts hold. 
The remainder of the paper describes this solution in detail.
\begin{comment}
In the
remainder of this paper we describe how we automatically translate
AGREE contracts into Simulink observer that can be used to formally verify the
original component artifacts.
\end{comment}

\subsection{Component Requirements and Verification}
\label{subsec:simulink}

The following tools and features are used to capture component level requirements and perform verification.  

\textbf{Simulink.\ } Simulink~\cite{Simulink}, developed by MathWorks and integrated
with MATLAB, provides a graphical programming environment for
modeling, simulation, code generation, testing, and analysis. It is
widely used in the Avionics industry. It is used to capture low-level
component design models and requirements.

\textbf{Simulink Design Verifier.\ } The Simulink Design Verifier (SLDV) tool \cite{SLDV},  provides a model checker for the Simulink environment.
SLDV can verify properties expressed with MATLAB functions, Simulink
blocks, or Stateflow diagrams. The first is a textual language while
the last two are graphical.

\textbf{Simulink observer.\ } A Simulink observer is a component in a
Simulink model which observes the behavior of another component and computes a Boolean value indicating if the latter component is satisfying its requirements. A Simulink observer along with the component it observes
can be verified using SLDV to show that the component under
observation always satisfies its requirements. Using DO-331 terminology, 
the Simulink observer is a specification model that captures high-level requirements, 
while the component it observes
is a design model that captures low-level requirements.  Our tool generates
Simulink observers using a MATLAB function block which encapsulates
a MATLAB function. A MATLAB function consists of statements written in
the MATLAB scripting language, an imperative, dynamically typed
language. In addition to a main function, a MATLAB function block can
contain other local functions defined in the same block. Unlike the
other graphical language alternatives, the textual representation of a
MATLAB function makes the export easier to control and maintain.

%%  LocalWords:  SLDV Stateflow

\section{Detailed Approach}
\label{sec:detailed_approach}

This section details our approach for automatically constructing a specification model from high level requirements.

\begin{table}
\caption{Mapping between AGREE and MATLAB Constructs}
\footnotesize
\begin{center}
\setlength{\tabcolsep}{0.5em} % for the horizontal padding
{\renewcommand{\arraystretch}{1.2}% for the vertical padding
\begin{tabular}{|c|c|}\hline
\textbf{AGREE Constructs} & \textbf{MATLAB Constructs}\\ \hline
Component contract\textsuperscript{1} & Simulink observer\\ \hline
Component inputs and outputs\textsuperscript{1} & Inputs to the
Simulink observer\textsuperscript{2} \\ \hline
Assume statement & Proof assumption \\
assume ``B input range'' : $\mathit{Input} < 20$ & sldv.assume$(\mathit{Input} < 20) $
\\ \hline
Guarantee statement & Proof objective\\
guarantee ``B output range'' : &
\multirow{2}{*}{sldv.prove$(\mathit{Output} < (\mathit{Input} +
  15))$}\\
$\mathit{Output} < \mathit{Input} + 15$ &  \\ \hline
Equation statement & Assignments \\
eq $\mathit{Active} : \mathit{bool} =
\mathit{not}\ \mathit{Sync.Active}$ & $\mathit{Active} =
\mathit{not}(\mathit{Sync.Active})$ \\\hline
If-then-else expression & Generated local function
\\
$\mathit{if}\ \mathit{Error}\ \mathit{then}\ \mathit{false}\ \mathit{else}\ \mathit{Active}$
& $\mathit{ifFunction}(\mathit{Error}, \mathit{false}, \mathit{Active})$ \\ \hline
AGREE basic data types & MATLAB built-in data types\textsuperscript{2} \\
int & (u)int8, (u)int16, (u)int32 \\
real    &  single, double \\
bool  & boolean \\ \hline
Record types (on inputs and outputs) & Simulink bus objects \\ \hline
AGREE operators\textsuperscript & MATLAB operators or function calls \\ 
-, not, $<>$, and, or & -, $\sim$, $\sim=$, \&\&, $\parallel$ \\
$+$, $-$, $*$, $/$, $>$, $<$, $>=$, $<=$ & $+$, $-$, $*$, $/$, $>$, $<$, $>=$, $<=$ \\
mod operator & mod function \\
= (equal operator) & isequal function\textsuperscript{3} \\
div (integer divide operator) & $/$ with operands typecast to integer types\\
$\Rightarrow$ & generated local $\mathit{impliesFunction}$\\
$\rightarrow$ & generated local $\mathit{arrowFunction}$\textsuperscript{4}\\
$pre$ & persistent variable for the
operand\textsuperscript{4}\\ \hline
\multicolumn{2}{l}{\textsuperscript{1} \footnotesize{This information
    comes from the component type in AADL.}}\\
\multicolumn{2}{l}{\textsuperscript{2} \footnotesize{Data size
    selection based on user input (Section
    \ref{subsec:data_types}).}}\\
\multicolumn{2}{l}{\textsuperscript{3} \footnotesize{Use isequal
    function rather than $==$ to apply to structure types}} \\
\multicolumn{2}{l}{\textsuperscript{4} \footnotesize{The translation
    for $\rightarrow$ and $pre$ operators need persistent variables
    (Section
    \ref{subsec:temporal_constructs}).}}\\
\end{tabular}
}
\end{center}
\label{table:translation_scheme}
\end{table}

\subsection{Export Scheme Overview}
\label{subsec:export_scheme}

The requirements used to generate each specification model come
directly from a component contract specified in AGREE. Each
specification model is a Simulink observer implemented as a MATLAB
function. The observer's interface is generated from the component's
features described in the AADL model.

Table~\ref{table:translation_scheme} provides a summary of the
constructs that appear in an AGREE contract and their mapping in
MATLAB. Our process can translate any AGREE specification. The specification model generation process 
is divided into two major steps, as depicted in Figure \ref{fig:impl-scheme}:

\begin{figure}
	\begin{center}
		\includegraphics [height=2.8in] {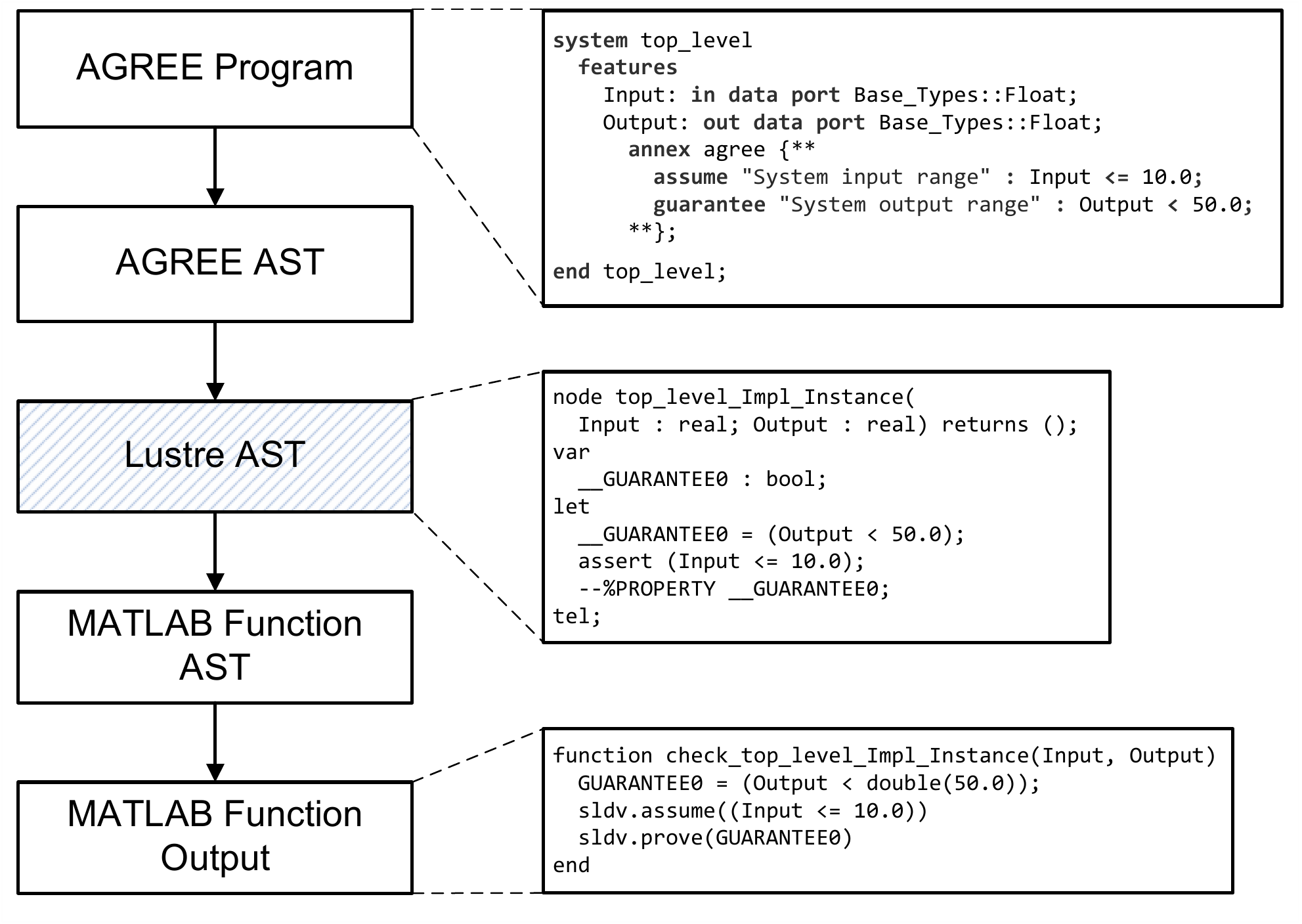}
	\end{center}
	\caption{Implementation Scheme}
	\label{fig:impl-scheme}
\end{figure}

\begin{enumerate} 
\item The tool produces an intermediate specification in Lustre. The
  Lustre language \cite{Lustre-91} is a synchronous dataflow language
  for modeling reactive systems, with formalisms similar to temporal
  logics \cite{Huth-Ryan}. The AGREE grammar and the Lustre grammar
  are very similar. This makes Lustre well suited as a common
  intermediate language to feed into different formal analysis or
  translation engines. A number of common translation steps are
  performed to create this intermediate format. For example, variable
  assignments are put into data-flow order, all function calls are
  inlined, and nested temporal expressions are decoupled.
	
\item From the intermediate Lustre a MATLAB function is produced. The
  MATLAB function is specified by an abstract syntax tree (AST). This
  allows for structured, easily extendable export. MATLAB specific
  features introduced in this translation include constructing valid
  MATLAB identifiers with no duplications and turning local
  structures in the intermediate output into local variables to
  eliminate any dynamically allocated structures.
\end{enumerate}

\subsection{Translation for Temporal Operations}
\label{subsec:temporal_constructs}

There are two types of temporal operations used in AGREE:

\begin{itemize} 
	\item The $\rightarrow$ operation evaluates to its left-hand side expression when the
	 transition system is in its initial state. Otherwise it evaluates to its right-hand side
	  expression. For example, the expression: $\mathit{true} \rightarrow \mathit{false}$ is
	   $\mathit{true}$ in the initial state and $\mathit{false}$ otherwise.
	
	\item The $\mathit{pre}$ operation takes a single expression as an argument and returns the
	 value of this expression in the previous state of the transition system. For example, the
	  expression: $x = (0 \rightarrow\mathit{pre}(x) + 1)$ constrains the current value of variable
	   $x$ to be 0 in the initial state; otherwise it is the value of $x$ in the previous state
	    incremented by 1. 

\end{itemize}

In the model's initial state the value of the $pre$ operation on any expression is undefined. 
Every occurrence of a $pre$ operator must be in a subexpression of the right hand side of the
 $\rightarrow$ operator. The $pre$ operation can be performed on expressions containing other $pre$
  operators, but there must be $\rightarrow$ operations between each occurrence of a $pre$ operation.
   For example, the expression: $true \rightarrow pre(pre(x))$ is not well-formed, but the expression: 
   $true \rightarrow pre(x \rightarrow pre(x))$ is well-formed.

To represent temporal constructs, the Simulink observer needs to
differentiate the behavior at the initial state from the other states.
It also needs to remember variable values from the previous calls to
the function. 

We make use of persistent variables to record the
previous state of the function's variables across multiple calls. A single persistent Boolean
 variable, $\mathit{first\_time}$, is used
for all $\rightarrow$ expressions to indicate whether or not the function is being called for the first time.
Additionally, a persistent variable is created for each unique
$\mathit{pre}$ expression\footnote{For example, if the term
  $\mathit{pre}(x)$ appears multiple times in the AGREE contract, we only
  create a single persistent variable for this expression.}. We refer
to these variables as the ``$\mathit{pre}$ variables''. The Simulink unit delay block could also
be used to remember previous variable values by placing the graphical block outside of the MATLAB
function for each ``$\mathit{pre}$ variable''. However, the block needs to be placed outside of the
MATLAB function, requiring any ``$\mathit{pre}$ variable'' to become an input to the function; 
the graphical representation also makes it harder to automate.

The persistent variables used to model the $\rightarrow$ and $pre$ operations appear
in the following contexts in the Simulink observer:

\begin{enumerate} 
	\item \textbf{Declaration.\ } Each of these persistent variables is
	declared at the beginning of the function. MATLAB is dynamically
	typed, so the type of these variables is determined during their
	initialization.
	\item \textbf{Initialization.\ } The initialization of a persistent
	variable occurs immediately after its declaration. The built-in
	function ``isempty'', e.g., 
	$\mathit{isempty}$ $(\mathit{first\_time})$, is used to determine whether or not the
	variable has been initialized.
	\begin{comment}
	For example, this is how the
	$\mathit{first\_time}$ variable is initialized:\\
	\vspace{-1em}
	\begin{center}
		\begin{minipage}{.3\textwidth}
			$\mathit{if}\ \mathit{isempty}(\mathit{first\_time})$\\
			\hspace*{1em}$\mathit{first\_time} = \mathit{true};$\\
			$\mathit{end}$\\
		\end{minipage}
	\end{center}
	\vspace{-1em}
	\end{comment}
	The $\mathit{pre}$ variables are initialized to the default value for
	their type (e.g., true for booleans, 0 for integers, and 0 for
	floating points). Because all occurrences of $\mathit{pre}$ operators
	are guarded by $\rightarrow$ operators, this initial value is never
	used. This initialization takes place for the sole purpose of allowing
	the Simulink code generator to function properly.
	
	\item \textbf{Use.\ } Each of these persistent variables is used in
	place of its corresponding $\mathit{pre}$ expression.
	
	\item \textbf{Update.\ } Before the observer function returns, all of
	the $\mathit{pre}$ variables are updated to the current value of
	their expression. For example, the persistent variable for the
	expression $\mathit{pre}(x)$ is updated to the value of $x$. 
	The $\mathit{first\_time}$ variable is always set to false
	before the observer function returns.
	
\end{enumerate}

\subsection{Translation for Data Types}
\label{subsec:data_types}
Here we note differences between the data types of AGREE and MATLAB.

\textbf{Constants.\ } Any constant numbers (integers or floats) that
appear in a MATLAB function are assumed to be $double$ precision
floating point numbers. Therefore, explicit typecasts are needed when
translating constants from the AGREE specifications.

\textbf{Arbitrary Data Size vs. Fixed Data Size.\ } AGREE assumes that
integer and real valued variables have arbitrary size. However,
MATLAB's primitive data types for integers are of bounded size
(integers are represented by 8-bit, 16-bit, or 32-bit 2's
complement numbers). Integer valued variables in AGREE are
translated into fixed size integers in MATLAB. Similarly, MATLAB uses 
floating point arithmetic to represent non-integers.  Real valued
variables in AGREE are translated to floating point variables in
MATLAB. The size/precision of the translated variables can be changed
easily by users.

This mismatch in types can cause differences in semantics for some
contracts described in AGREE and their corresponding Simulink observers.
Moreover, we note that SLDV interprets floating point variables as real
variables as well. So it suffers from the same mismatch in semantics
for floating point vs. real numbers. In the future we plan to allow
users to specify bit-vector types in AGREE.
 
\subsection{Workflow}
\label{subsec:impl}

We have implemented the export scheme as an extension
to AGREE, available at~\cite{SMACCM-Toolset}. 
The export process has the following steps:
\begin{enumerate} 
\item \textbf{Select Data Types.\ } Users may select one of the
  MATLAB/Simulink supported integer types 
  %\footnote{MATLAB supports (u)int64 but Simulink does not, so it 
  %	is excluded from the options.} 
  (i.e., (u)int8, (u)int16, (u)int32) 
  to represent integers from the AGREE specification and one of the MATLAB
  supported floating point types (i.e., single, double) to represent
  reals from the AGREE specification.

\item \textbf{Export Design Contracts.\ } For any component with an
  AGREE contract, the user can invoke the tool to translate the contract
  into a MATLAB function.

\item \textbf{Update Simulink Model.\ } A script file
  provided by the toolset automatically packages the MATLAB function
  generated above as a MATLAB function block and connects the
  function block to the inputs and outputs of the component's Simulink
  model. The augmented Simulink model 
  %is the verification model for the component that 
  contains both a design model and a specification model
  that observes the design model.

\item \textbf{Invoke Simulink Design Verifier (SLDV).\ } Users can
  invoke SLDV on the verification model generated in the above step.
  SLDV checks to see if all properties in the MATLAB function are
  true, and provides counterexamples for the ones that are
  falsified.
\end{enumerate}

%%  LocalWords:  sldv eq bool ifFunction isequal impliesFunction
%%  LocalWords:  arrowFunction dataflow formalisms isempty
%%  LocalWords:  Booleans

\section{Case Studies}
\label{sec:case_study}

In this section, we evaluate and demonstrate the effectiveness of the
export techniques in two case studies: 1) an avionics braking and
steering control unit and 2) a medical infusion pump. The workflow was 
tested with the latest version of the AGREE toolset and MATLAB Release 2015b.

As our export tool has not yet been qualified~\cite{DO-330}, for each
case study, the specification model generated by the tool is manually
reviewed against the original contracts and design information in
AGREE and AADL to assess if the high-level requirements have been
maintained.

\subsection{Avionics Braking and Steering Control Unit} 

\textbf{Overview.\ } The avionics Braking and Steering Control Unit
(BSCU) is a computer located in an aircraft's Wheel Braking System
(WBS), controlling the ``Normal braking, Autobrake, Nose Wheel
Steering Aid and Antiskid functions''~\cite{Quasi-Synchronous}. The
specification of the BSCU came from prior verification
efforts~\cite{Quasi-Synchronous} based on the report of an Airbus
A-320 accident which occurred on May 21,
1998~\cite{BSCU-accident-report}. In that accident, both the normal
and alternate braking systems failed on landing. The loss of the
normal braking system was caused by logic disagreement in the BSCU.

The BSCU system consists of two functionally identical channels, with
only one channel being active at a time. When a fault is detected in
the active channel, the standby channel becomes active if it is not
faulty. Each channel contains a command function unit (COM component)
and a monitor function unit (MON component). Both the COM and MON
components compute the braking pressure to be applied based on their
braking mode. Their outputs are compared at the MON function unit, and
a fault will be logged when there is a disagreement between the
outputs.

The COM and MON units operate in four braking modes: MANUAL, LO, MED,
MAX. In MANUAL mode, the computed breaking pressure is mainly
determined by the pressure on the brake pedal applied by the pilot.
Other modes are Autobrake modes selected when pilot presses one of the
LO, MED, or MAX buttons on the AUTO BRK panel, providing low, medium,
and maximum levels of deceleration. Each unit starts in the MANUAL
mode, and can transition to another mode when the associated button is
pressed once; pressing the same button again transitions the unit back
to MANUAL mode.

For this specific case study, the system architecture was previously
modeled in AADL, and the design contracts between the components were
specified in AGREE.  
Prior work~\cite{Quasi-Synchronous} has found a
disagreement in detecting a button push between the COM and MON
component. The problem was remedied by updating the design contracts in the
architecture model.

For this case study, we created Simulink models for the COM and MON
components. This was a manual design process to interpret the
high-level requirements into a design model. The behaviors of the COM
and MON models were intended to satisfy all AGREE contracts for the
COM and MON components in AADL.

For each component, we exported the design contract to a Simulink
observer and connected it to the corresponding Simulink model. 
%The augmented model encompassed both the design model and the
%specification model for the component. 
We ran Simulink Design
Verifier (SLDV) on the augmented model to discover which properties were
validated and which were not. For the falsified properties, SLDV
produced a counterexample.

\textbf{Results and Findings.\ } Two types of falsified properties
were found during the verification process. The first type was caused
by a discrepancy on the behavior of the initial step. The second type
was due to a discrepancy on the value of a global parameter used to
indicate if the component is currently in an active channel. In both
cases, the Simulink design model failed to interpret the specific
design detail as presented in AGREE. Such discrepancies were missed
from the first round of manual review of the models. After updating
the Simulink model to match the design contracts, all properties
specified in the Simulink observer were verified.

Investigation of the counterexamples was carried out by comparing the
values of the intermediate signals computed in the model and in the
Simulink observer during the simulation of the counterexamples.
Having the specification model and design model co-located in the same
environment allowed the simulation to compare their values during
runtime.

The verification results demonstrate the benefit of using formal
verification over manual review or simulation/testing, as it reasons
about all execution paths and identifies design flaws that can be
missed by other methods.

Automatically exporting AGREE contracts to Simulink observers allows
fast turn-around in verification. The verification of the Simulink
design model can be conducted as soon as the model is created. This
supports early and frequent verification starting from the design phase.
It also reduces errors that are easily introduced from manual
interpretations, especially for large components with complex
contracts.

%%  LocalWords:  BSCU WBS Autobrake Antiskid BRK AADL Simulink SLDV

\subsection{Generic Patient Controlled Analgesic Infusion Software}

\textbf{Overview.\ } The Generic Patient Controlled Analgesic (GPCA)
infusion pump system~\cite{Murugesan:2013:CVM:GPCA} is a medical
cyber-physical system ``used for controlled delivery of liquid drugs
into a patient's body according to a physician's prescription (the set
of instructions that governs infusion rates for a
medication).''~\cite{Murugesan:2013:CVM:GPCA}. It allows patients to
administer a controlled amount of drug (typically a pain medication)
themselves. It consists of four main components: Alarm, Infusion,
Mode, and Logging. They are used to monitor the exceptional conditions
and notify the clinician, determine the flow of drug, manage the mode,
and log the status of the system. Detailed information on GPCA
requirements can be found in~\cite{GPCA-Project-UMN}.

The workflow for this case study was similar to the BSCU case study,
except that the Simulink design models for the components, as well as
the Simulink observer for the properties of the design models, had
been manually created in prior work~\cite{GPCA-Project-UMN}. In this
case study, we reused the design model created for each component, and
we replaced the existing (manually created) Simulink observers with the
ones generated by our tool from the corresponding AGREE contracts. The
updated models were then verified using Simulink Design Verifier and
the verification results from both models were compared.

\textbf{Results and Findings.\ } For both the manual and auto version
of the Simulink observer, the verification results identified
falsified properties due to the design model not behaving as expected.
The verification time between the two versions were comparable
(within 10 seconds). Some properties were undecided after
reaching the maximum analysis time (set at 1200 seconds) for both
versions.

Although the manually created versions of the Simulink observers are still a work
in progress, we can make the following observations:

\begin{enumerate} 
\item The manual properties tended to address the simpler, more
  straightforward contracts in AGREE, and they often missed modeling
  the temporal constructs from AGREE (i.e.,$\mathit{pre}$ and
  $\rightarrow$ operators).  Automation now allows us to easily translate
  even the complex contracts.  
\item The manual properties tended to lag behind the AGREE contract
  updates, resulting in different verification results between the
  manual and automated versions for the same AGREE contract.  Automation
  makes it easy to keep all the models synchronized.  
\item The manual properties used Simulink unit delays outside of the
  MATLAB function to interpret the $\mathit{pre}$ operator, a
  translation that preserves the meaning but is not easy to automate.
\item The manual properties selected signals from bus elements outside
  of the MATLAB function, while the auto translated properties did bus
  element selection inside the MATLAB function. The latter is a design
  choice that is easier to automate and maintain.
\end{enumerate}

We found the benefits of automatically connecting the created Simulink observer
to the design model through a counterexample. In this
counterexample, one input port to the design model and the Simulink
observer was a duplicate (different port numbers and treated as
different ports) instead of a replicate (same port number and treated
as the same port). This made the observer not a synchronous one, and
yielded different verification results from the version that had the
Simulink observer auto connected.

We also found design details introduced in the Simulink model that did
not conform to the interface design in AGREE. For example, for an
input port of record type in AGREE, its counterpart in Simulink
(of Simulink bus type) had additional elements and elements with
different names. While it is understandable that the design model may
introduce new details needed for the component, any new design details
that affect the interface should be synchronized with the AGREE model.

\section{Related Work}
\label{sec:related_work}

The idea of auto generating test cases from higher level requirements
has been the subject of intensive study in both the academia and
industry~\cite{Wang:2015:Test-case-gen-from-use-case-spec,Ibrahim:2007:Gen-test-case-from-UML,Escalona:2011:Survey-on-auto-test-gen}.
Creating properties for formal verification from higher level
requirements, has been performed
manually~\cite{Miller:2006:Proving-the-shalls,Murugesan:2013:CVM:GPCA}, through
patterns~\cite{Bozzano-Cimatti-COMPASS-14}, and
automatically~\cite{Silva-13-Property-Gen,Soeken-11-Property-Gen}. The
unique contribution of our work is a method for automatically
exporting high-level requirements from a system-level reasoning
framework as property observers in a component-level modeling framework.
This enables formal verification of the component requirements as they
are developed, bridging the gap between system-level and
component-level reasoning. The compositional reasoning framework
OCRA~\cite{OCRA} has similar goals as AGREE. Both frameworks reason
hierarchically about a system of components with connections and
contracts. However, as far as we know, there are no tools to translate
OCRA contracts to observers in specification languages commonly used
in the avionics industry.

\section{Conclusions}
\label{sec:conclusion}

In this paper we have described a method for translating design
contracts for components in an AADL software architecture model into specification
models that can be verified at the component level.  We have provided tool support for
export as Simulink observers that can be verified using the Simulink Design Verifier.  Moreover, 
our approach is sufficiently general that other component development environments could be
easily targeted.  This approach is built upon the AGREE compositional analysis framework that allows verification
of requirements during architecture development, prior to software component implementation.
Applying the technique on an avionics system and a medical device system has
shown that the design contracts from the architecture model were faithfully
exported, and saved time and reduced errors compared to the manual effort.  Our approach also
allowed verification to proceed in parallel with software development.

Possible future work includes qualifying the export tool in accordance with avionics
certification guidelines~\cite{DO-330} and enhancing the usability of the tool
by supporting automatic re-verification when design contracts are updated.

\subsubsection*{Acknowledgments.} This work was funded by NASA under
contract\\ NNA13AA21C (Compositional Verification of Flight Critical
Systems). We would like to thank Chad Van Fleet, Anitha
Murugesan, and Mike Whalen for their valuable feedback during this
work.

%%  LocalWords:  NNA Anitha Murugesan Whalen

\bibliographystyle{splncs}
\bibliography{references} 

\end{document}